\documentclass[]{aastex631}

\newcommand{\fo}{\ensuremath{f^\parallel}}
\newcommand{\fe}{\ensuremath{f^\perp}}
\newcommand{\ord}{parallel}
\newcommand{\xord}{perpendicular}
\newcommand{\pq}{\ensuremath{P_Q}}
\newcommand{\pu}{\ensuremath{P_U}}
\newcommand{\px}{\ensuremath{P_X}}
\newcommand{\pr}{\ensuremath{P_{\rm r}}}
\newcommand{\nnq}{\ensuremath{N_Q}}
\newcommand{\nnu}{\ensuremath{N_U}}
\newcommand{\nnx}{\ensuremath{N_X}}

\newcommand{\prn}{\ensuremath{p_{\rm r}}}

\begin{document}

\title{Optical spectropolarimetry of binary asteroid Didymos-Dimorphos before and after the DART impact}

\author[0000-0002-7156-8029]{Stefano Bagnulo}
\affiliation{Armagh Observatory \& Planetarium, College Hill, Armagh, BT61 9DG, UK}

\author[0000-0002-6610-1897]{Zuri Gray}
\affiliation{Armagh Observatory \& Planetarium, College Hill, Armagh, BT61 9DG, UK}
\affiliation{Mullard Space Science Laboratory, University College London, Holmbury St. Mary, Dorking RH5 6NT, UK}

\author[0000-0002-5624-1888]{Mikael Granvik}
\affiliation{Department of Physics, PO Box 64, FI-00014 University of Helsinki, Finland}
\affiliation{Asteroid Engineering Laboratory, Lule\r{a} University of Technology, Box 848, SE-98128 Kiruna, Sweden}

\author[0000-0002-6645-334X]{Alberto Cellino}
\affiliation{INAF -- Osservatorio Astrofisico di Torino, I-10025 Pino Torinese, Italy}

\author[0000-0002-9321-3202]{Ludmilla Kolokolova}
\affiliation{Department of Astronomy, University of Maryland, College Park, MD 20742-2421, US}

\author[0000-0001-8058-2642]{Karri Muinonen}
\affiliation{Department of Physics, PO Box 64, FI-00014 University of Helsinki, Finland}

\author[0000-0002-5138-3932]{Olga Mu\~{n}oz}
\affiliation{Instituto de Astrof\'{\i}sica de Andaluc\'{\i}a, CSIC, Glorieta de la Astronomia s/n, E-18008 Granada, Spain\\}

\author[0000-0002-9298-7484]{Cyrielle Opitom}
\affiliation{Institute for Astronomy, University of Edinburgh, Royal Observatory, Edinburgh, EH9 3HJ, UK}

\author[0000-0001-7403-1721]{Antti Penttil\"a}
\affiliation{Department of Physics, PO Box 64, FI-00014 University of Helsinki, Finland}

\author[0000-0001-9328-2905]{Colin Snodgrass}
\affiliation{Institute for Astronomy, University of Edinburgh, Royal Observatory, Edinburgh, EH9 3HJ, UK}

\begin{abstract}
We have monitored the Didymos-Dimorphos binary asteroid in spectropolarimetric mode in the optical range before and after the DART impact. The ultimate goal was to obtain constraints on the characteristics of the ejected dust for modelling purposes. Before impact, Didymos exhibited a linear polarization rapidly increasing with phase angle, reaching a level of $\sim 5$\,\% in the blue and $\sim 4.5$\,\% in the red. The shape of the polarization spectrum was anti-correlated with that of its reflectance spectrum, which appeared typical of an S-class asteroid. After impact, the level of polarization dropped by about 1 percentage point (pp) in the blue band and about 0.5\,pp in the red band, then continued to linearly increase with phase angle, with a slope similar to that measured prior to impact. The polarization spectra, once normalised by their values at an arbitrary wavelength, show very little or no change over the course of all observations, before and after impact.  The lack of any remarkable change in the shape of the polarization spectrum after impact suggests that the way in which polarization varies with wavelength depends on the composition of the scattering material, rather than on its structure, be this a surface or a debris cloud. 
\end{abstract}


\section{Introduction}
\label{Intro}
The goal of NASA's Double Asteroid Redirection Test (DART) was to test the deflection of a near-Earth object (NEO) through an impact by a dedicated spacecraft in the context of planetary defense \citep{DART-summary-paper}. DART was launched on November 23, 2021 for a 10-month journey to the binary asteroid Didymos. The targeted system is composed of the 780-m asteroid Didymos (primary) around which orbits the 160-m secondary Dimorphos. The DART probe impacted Dimorphos head-on on September 26, 2022 at 23:14 UT, changing its orbital period around the primary by about 30\,min and enabling the measurement of the efficiency of the kinetic impactor technique in producing a change in linear momentum of the target \citep{Cristinas-paper-on-period-change}.
The epoch of the impact was chosen to optimize the geometry for ground-based observing, hoping that observations will allow the assessment of the excess linear momentum exerted on Dimorphos as a result of ejecta leaving its surface due to the DART impact --- which is essential for extrapolating the results of the DART impact on Dimorphos to other objects.

Linear polarization measurements have long been used as a remote sensing tool for the characterisation of objects in the Solar System. Measurements in the standard optical filters may be plotted as a function of the phase angle $\alpha$, that is, the angle between the Sun and the observer, as seen from the target object. 
A simple Rayleigh scattering model predicts that the plane of polarization should be perpendicular to the scattering plane (the plane determined by the Sun, the observer, and the target object). In fact, at small phase angles ($\alpha \la 20^\circ$), the polarization is oriented in the direction parallel to the scattering plane --- this phenomenon of {\it negative polarization} is commonly interpreted in terms of coherent backscattering \citep[e.g.;][and references therein]{Muietal02,Muinonen04} and scattering by wavelength-scale particles  \citep[e.g.; ][]{LummeRahola98,Muinonenetal11}. At larger phase angles, the polarization increases up to $\alpha \sim 90$--$100^\circ$, although main-belt asteroids may be observed only up to a maximum of $\alpha \simeq 30$--$35^\circ$, where the polarization reaches at most a value of 1--2\,\%. The analysis of global shape of the phase-polarization curves, combined with photometric and spectroscopic data, provides a way to assess the physical properties of the topmost surface layer of the surfaces of small Solar System bodies, including the complex refractive index, particle size, packing density, and microscopic optical heterogeneity.

Important constraints may also be provided by the observations of how polarization depends upon wavelength. While multi-band polarimetric measurements have been obtained and analysed in the past \citep[e.g.;][]{Belskayaetal09}, spectropolarimetry of asteroids is still a relatively unexploited technique; to our best knowledge, the only dataset published in the literature so far is the outcome of a mini-survey by \citet{Bagetal15}, who have shown, for instance, that objects characterized by very similar reflectance spectra may exhibit quite different polarized spectra, underscoring the fact that intensity measurements alone may not be sufficient to fully characterise the light scattering properties of planetary bodies. We therefore decided that 
a multi-wavelength polarimetric monitoring of the Didymos-Dimorphos binary system, before and after the DART impact, would be especially useful to characterize the possible changes in the polarimetric properties of a binary system, as a consequence of the excavation and ejection of a sub-surface layer of regolith, and subsequent variation of surface properties of the two components of the binary system, due to ejecta impacts on Didymos or Dimorphos. These observations are unique also because they are obtained at phase angles not observable from Earth for main-belt asteroids -- only an handful of NEOs has been observed with broadband polarimetric techniques in the past.

In this letter, we describe spectropolarimetric measurements obtained at the ESO VLT, before and after impact, and provide a preliminary assessment of what these measurements tell us about the ejecta properties. 

\section{Observations}\label{Sect_Observations}

Polarimetry of the Didymos-Dimorphos system was acquired before and after the DART impact with the FORS2 instrument on the ESO VLT both in spectropolarimetric and imaging polarimetric mode, as well as through aperture polarimetry with the ALFOSC instrument on the Nordic Optical Telescope (NOT). The observing campaign at the NOT is still ongoing. Here we report the 10 polarization spectra obtained with VLT FORS2 from August 23 to October 24, five prior to the DART impact and five after the DART impact. The quantity we are interested in as a function of the phase angle is
\begin{equation}
\pr (\lambda)= \frac{F_\perp - F_\parallel}{F_\perp + F_\parallel}
\label{Eq_Pr}
\end{equation}
where $F_\parallel$ and $F_\perp$ refer to the fluxes measured through a linear polarizer parallel or perpendicular to the scattering plane, respectively. The definition of Eq.~(\ref{Eq_Pr}) is equivalent to that given by \citet{Zeletal74} as $\pr = P \cos(2\theta_{\rm r})$, where $P$ is the observed fraction of linear polarization, and $\theta_{\rm r}$ is the angle between the position angle of the observed polarization and the normal to the plane of the scattering. Technical details about observing strategy, data reduction, and quality checks are given in Appendix~\ref{Appendix}.

\section{Results}\label{Sect_Results}
\begin{figure*}
\begin{center}
\includegraphics[angle=0,width=12.0cm,trim={1.0cm 2.5cm 1.0cm 0.5cm},clip]{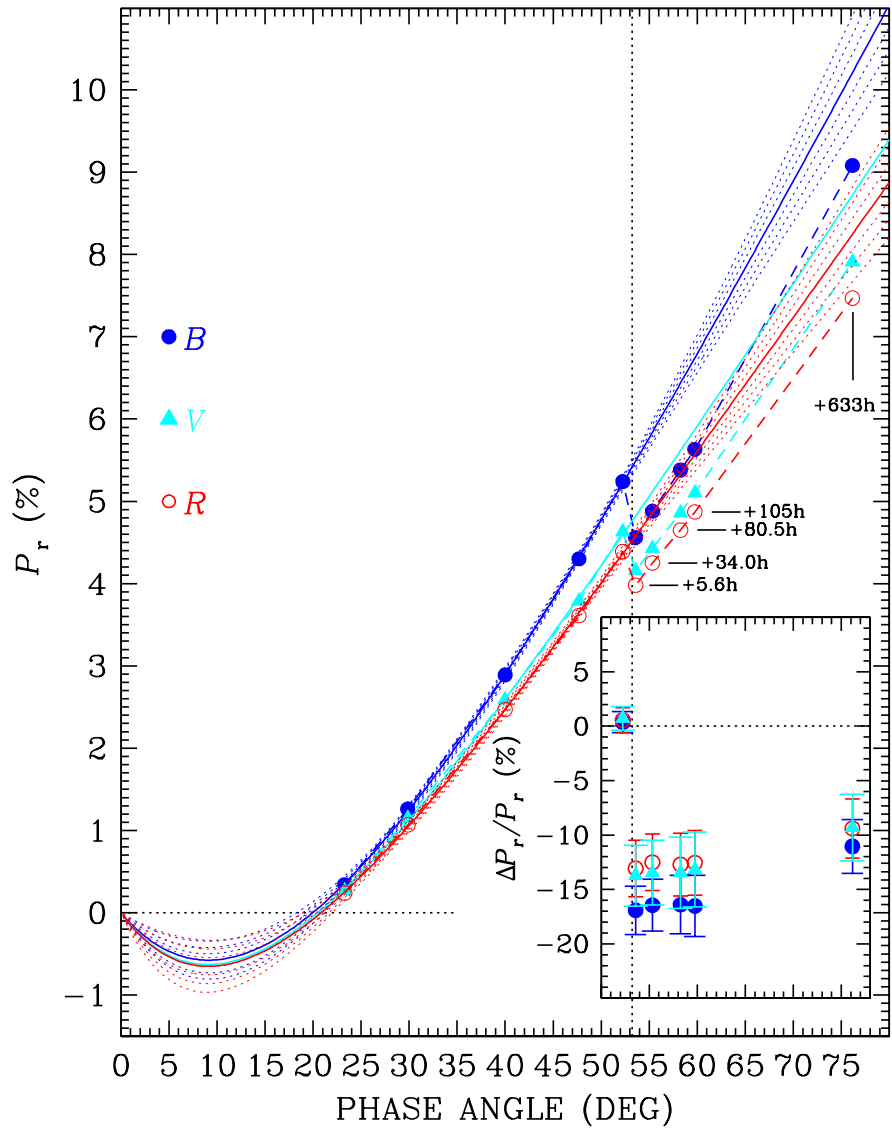}
\end{center}
\caption{\label{Fig_Pcurve} Solid circles joined by dashed curves show the synthetic broadband polarization values as a function of the phase angle obtained from FORS2 spectropolarimetric observations. Solid curves represent the best fits obtained with Eq.~(\ref{Eq_Fit}) on the data obtained prior to impact. Different symbols and colors refer to different filters as shown in the legend. The errorbars would have a size about symbol size or smaller, hence are not shown. The dotted lines represent the $\pm 1\,\sigma$ uncertainties around the best-fits on the $B$ and $R$ filter values. The vertical dotted line shows the phase angle at the time of impact. Data for the $I$ filter are not plotted as they would largely overlap with the data in the $R$ filter. Post-impact data are marked with the time at which they were obtained measured in hours from impact. The inset shows the fractional variation of the observed polarization as explained in the text.} 
\end{figure*}
The values of the synthetic broadband polarization measurements in the \emph{BVRI} filters as a function of phase angle, calculated as explained in Sect.~\ref{Sect_BBLP}, are given in Table~\ref{Tab_BBLP} and plotted in Fig.~\ref{Fig_Pcurve}. It appears that in the filters \emph{BVRI} the polarization decreases with wavelength, reaching a plateau around the $R$ and $I$ filter (the data for the $I$ filter are not shown in Fig.~\ref{Fig_Pcurve} as they would overlap with the data in $R$).  As will be further discussed in Sect.~\ref{Sect_Discussion}, polarization spectra extend to the very near IR, and contain information also at wavelengths longer than those of the optical filters. The spectra show that at wavelengths longer than the $I$ filter the polarization starts to increase with wavelength, reaching values similar to those of the $V$ band at $\lambda \sim 900$\,nm.

Figure~\ref{Fig_Pcurve} shows that the DART impact produced a dramatic effect: immediately after the impact the polarization dropped by $\sim 1$ percentage point (pp) in the $B$ filter and $\sim 0.5$\,pp in the $R$ filter.  
After the polarization drop induced by the impact, linear polarization continued to increase with phase angle with a slope similar to that observed prior to the impact.  

To compare the polarization observed post-impact with the values that would have been expected from the untouched Didymos-Dimorphos system, it is useful to fit the pre-impact polarimetric data with 
the linear-exponential function
\begin{equation}
  \pr(\alpha) = A \left({\rm e}^{-(\alpha/B)} - 1\right) + C\,\alpha
\label{Eq_Fit}
\end{equation}
in which $\alpha$ is the phase angle, and $A$, $B$, and $C$ are free parameters. This
semi-empirical function \citep{Kaaetal03} was found suitable to 
fit phase-polarization curves in asteroid polarimetry, at least up to phase angles $\sim 50^\circ$
\citep[see also][and references therein]{Celetal15, Beletal17, CelBennu}. In fact, the trigonometric function
\begin{equation}
  \pr(\alpha) =  b\,\sin^{c_1}\!\alpha \; \cos^{c_2}\! {\textstyle \frac{1}{2}} \alpha \;\sin(\alpha-\alpha_0) 
\label{Eq_For}
\end{equation}
in which $b$, $c_1$, $c_2$, and $\alpha_0$ are free parameters, would potentially provide a suitable description also at much larger phase angles \citep[e.g.;][]{Penetal05}. However, with only five pre-impact observations for each filter obtained at phase angles between $\sim24^\circ$ and $52^\circ$, the free parameters are not sufficiently constrained to allow for an accurate extrapolation. 

The best-fitting curves for the measurements obtained in the various filters using Eq.~(\ref{Eq_Fit}) are shown with solid lines in Fig.~\ref{Fig_Pcurve}. For the cases of the $B$ and $R$ filters, Fig.~\ref{Fig_Pcurve} also shows the uncertainties ($\pm 1, \pm 2, \pm 3$-$\sigma$ intervals) of the fitted solutions. 

The small inset of Fig.~\ref{Fig_Pcurve} shows the quantity
\begin{equation}
100\,\frac{\Delta \pr}{\pr} = 100\,\frac{\pr(\alpha) - \pr^{\rm (fit)}(\alpha)}{\pr^{\rm (fit)}(\alpha)}
\end{equation}
where \pr\ is the observed broadband polarization, and $\pr^{\rm (fit)}$ the value predicted by the best-fit solution based on the pre-impact data.  The result of this exercise convincingly shows that, for several days after impact, the polarization of the system was not back to values close to those that would have been expected without impact, and that the relative variation of the polarization was more pronounced at shorter than at longer wavelengths. Figure~\ref{Fig_Pcurve} suggests also that this lower polarization level persisted even until at least 4 weeks after impact. However, this latter result crucially relies on the assumption that the exponential function of Eq.~(\ref{Eq_Fit}) provides a realistic representation of the polarimetric curve even at phase angles as large as $\sim 75^\circ$, therefore it should be taken with caution.
\begin{table}
\caption{\label{Tab_BBLP} Broadband polarization values in the $BVRI$ filters measured from the spectra. Column~2 gives the time from impact. The uncertainties on the polarization values, dominated by systematics and not by photon-noise, are estimated of the order of 0.05\,\%.} 
\begin{center}
\begin{tabular}{crcrrrr}
\hline\hline
Date       & \multicolumn{1}{c}{$\Delta t$} & $\alpha$& \multicolumn{4}{c}{\pr (\%)} \\
YYYY-MM-DD &            & (deg)   & $B$\ \ & $V$\ \ & $R$\ \ & $I$\ \ \\
\hline
2022-08-28 &  $-$716h 25m & 23.29 & 0.34 & 0.27 & 0.24 & 0.19 \\  
2022-09-07 &  $-$471h 44m & 29.86 & 1.26 & 1.15 & 1.07 & 1.04 \\    
2022-09-17 &  $-$232h 54m & 39.99 & 2.89 & 2.59 & 2.47 & 2.46 \\   
2022-09-23 &\ $-$91h  49m & 47.68 & 4.30 & 3.79 & 3.61 & 3.59 \\   
2022-09-26 &\ $-$15h  49m & 52.25 & 5.24 & 4.63 & 4.39 & 4.35 \\ [2mm]
2022-09-27 &\ $+$05h  37m & 53.58 & 4.56 & 4.16 & 3.98 & 4.01 \\
2022-09-28 &\ $+$33h  59m & 55.34 & 4.88 & 4.43 & 4.25 & 4.27 \\
2022-09-30 &  $+$80h  30m & 58.26 & 5.39 & 4.86 & 4.65 & 4.65 \\
2022-10-01 &  $+$104h 59m & 59.76 & 5.63 & 5.10 & 4.87 & 4.84 \\
2022-10-24 &  $+$633h 01m & 76.16 & 9.08 & 7.91 & 7.47 & 7.37 \\
\hline
\end{tabular}
\end{center}
\end{table}

\begin{figure}
\begin{center}
\includegraphics[angle=0,width=7.0cm,trim={0.5cm 2.4cm 0.5cm 0.5cm},clip]{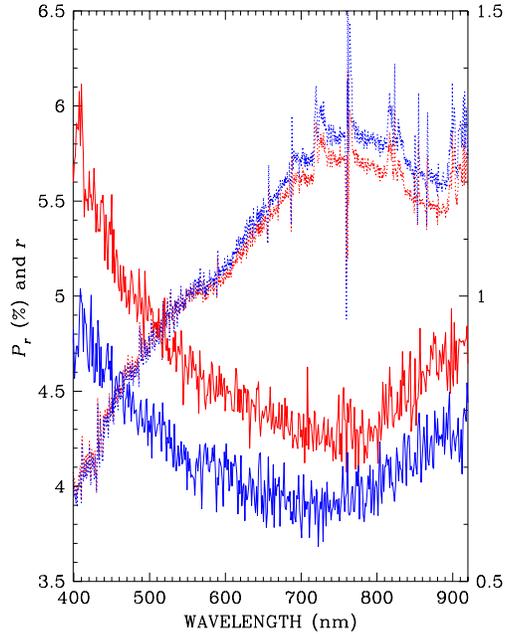}\ \
\end{center}
\caption{\label{Fig_Two_Spectra} Polarization spectra obtained with FORS2 at the phase angles before (red solid line) and after impact (blue solid line), with scale on the left axis. Dotted lines show the approximate relative reflectance spectra (normalized to 1 at 550~nm), with scale on the right axis.}
\end{figure}

\begin{figure*}
\begin{center}
\includegraphics[angle=0,width=18.0cm,trim={1.5cm 2.5cm 1.3cm 0.5cm},clip]{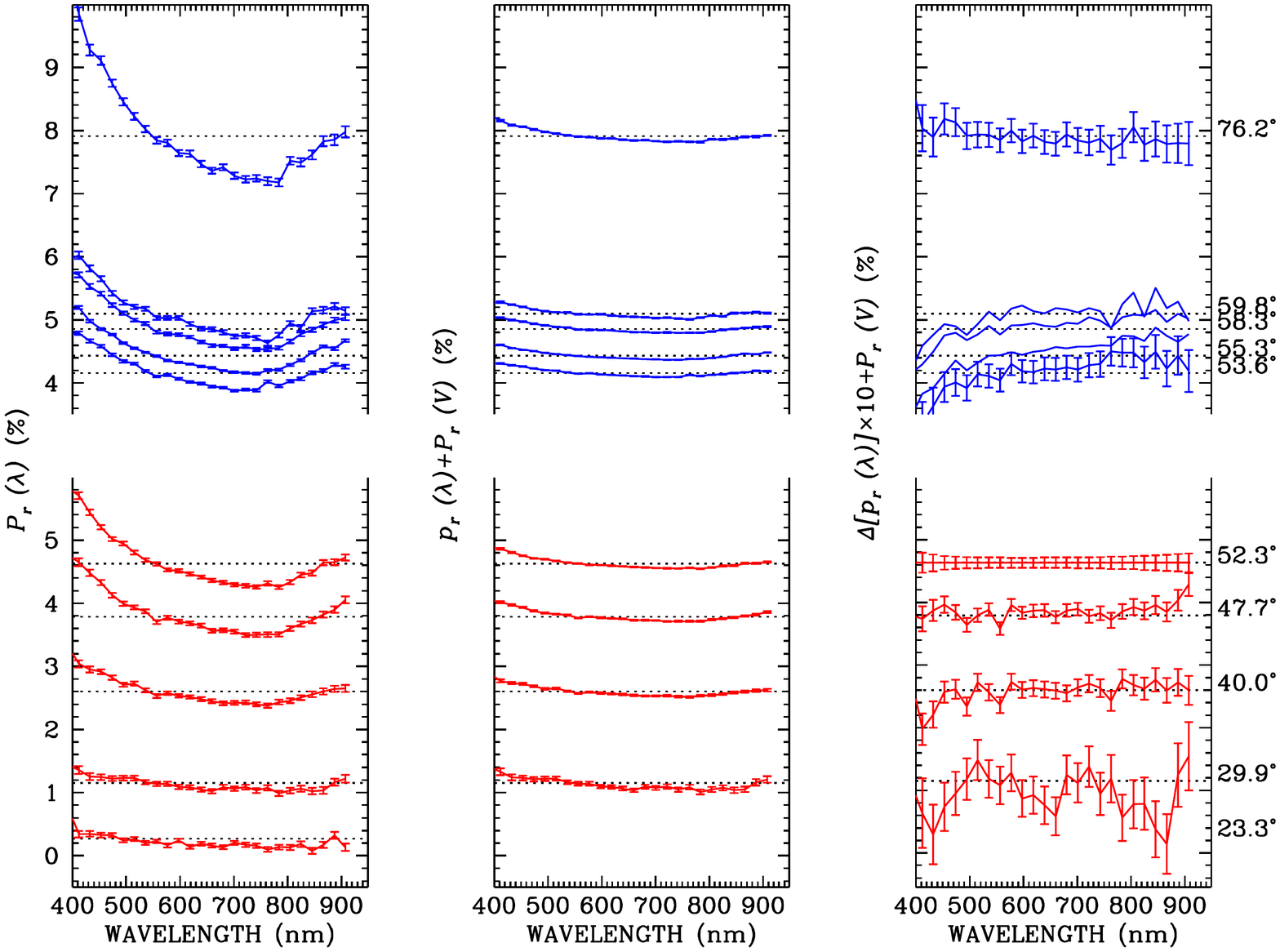}
\end{center}
\caption{\label{Fig_All_Spectra} Spectropolarimetry obtained at the phase angles shown on the right hand side of the plots. The bottom panels refer to data obtained prior to impact, the upper panels to data after the impact. The leftmost panels show the polarization spectra $\pr(\lambda)$. The mid-panels show the same spectra normalised to the values in the $V$ filter. For display purposes, they are vertically offset by a quantity equal to the normalisation factor. The right panels show the difference between each normalised polarization spectrum and the one obtained the night before the impact (26 September at 07:26 UT) multiplied by 10. Again, for display purposes, these data have been offset by the normalisation quantity $\pr(V)$.  For clarity, among the group of post-impact spectra (upper panel), error bars are shown only for one representative case.
In the middle and right panels, we have omitted to represent the polarization spectrum obtained on 23 August 2022, because, being taken close to the inversion angle, it would represent a noisy ratio between two small quantities.}
\end{figure*}

Figure~\ref{Fig_Two_Spectra} shows the reflectance and the polarization spectra obtained before the DART impact (26 September at 07:26 UT) and after the impact (27 September at 04:52 UT).
One can clearly see the drop of polarization described by the wavelength-integrated values of Table~\ref{Tab_BBLP}, while no obvious change of the overall shape of the polarization spectrum as a function of wavelength is evident. Both polarization spectra are inversely correlated with the reflectance spectra, as predicted by the Umov law \citep{Umov05} and reflectance (shown with dotted lines) seems slightly increased at longer wavelengths compared to shorter wavelengths. However, it must be clearly remarked that our spectrophotometric calibration (described in Sect.~\ref{Sect_Reflectance}) is very approximate, and the dotted lines of Fig~\ref{Fig_Two_Spectra} should serve the only purpose to verify that the observed polarization obeys the Umov's law, something that does not always happen with asteroid polarization spectra \citep{Bagetal15}.

The left panel of Fig.~\ref{Fig_All_Spectra} shows all polarization spectra obtained at various phase angles. Shape and amplitude of the polarization spectra depend on the phase angle and on wavelength. Following \citet{Bagetal15}, we assume that the dependence of the polarization upon phase angle may be separated from the dependence upon wavelength as the product of two functions $\pr(\lambda,\alpha) = P(\lambda) \ \mathcal{A}(\alpha)$. The polarization spectra normalised to the value measured at an arbitrary wavelength $\lambda = \lambda_0$  
\begin{equation}
\prn(\lambda)=\frac{\pr (\lambda,\alpha)}{\pr(\lambda=\lambda_0,\alpha)}  = 
\frac{P(\lambda) \mathcal{A}(\alpha)}{P(\lambda=\lambda_0) \mathcal{A}(\alpha)} 
\label{Eq_PNorm}
\end{equation}
are therefore independent of phase angle. The middle panels of Fig.~\ref{Fig_All_Spectra} show the normalised polarization spectra defined by Eq.~(\ref{Eq_PNorm}), which allow us to better compare the shape of the spectra among themselves (each spectrum was normalised to the value in the $V$ band). The rightmost panel shows the difference between all normalised polarization spectra and the one obtained on September 26, about 20\,h before impact, multiplied by 10. It is possible to appreciate a small change produced by the impact, in that, at shorter wavelengths, the post-impact neutralization of polarization seems marginally more efficient in the blue than in the red. This change could be caused by some instrumental effect, but since it is present in all spectra after the impact, and in none of the spectra obtained prior to the impact, it is possible that this tiny change is real, if marginal. According the the Umov law, this change in polarization should come with a relative increase of the reflectance at shorter wavelengths and decrease of the reflectance at longer wavelegnths. This is not confirmed by our FORS2 reflectance spectra, but it must be strongly remarked that our spectrophotometric calibration (described in Sect.~\ref{Sect_Reflectance}) is only approximate and cannot really be used to validate the results from polarimetry.

\section{Discussion}\label{Sect_Discussion}
Our observations represent the first spectropolarimetric monitoring of a binary asteroid system (the single components are too small and close to be resolved) over a large phase-angle range (20 to $75^\circ$). Most important, our observations of the DART mission offer a unique opportunity to compare the polarimetric properties of the light scattered by a surface (pre-impact) and by a dust cloud (presumably) of similar composition (post-impact). Previous potential examples of similar polarimetric observations could be those of cometary nuclei \citep[e.g., 2P/Encke,][]{Boeetal08} or those of ‘main-belt comets’, or ‘active asteroids’ \citep[e.g., 133P/Elst-Pizarro or P/2010 R2,][]{Bagetal10,Moreno2011ApJ}, which display a coma and/or tail close to perihelion, and offer the opportunity to study their surface directly as well as their ejecta. Despite the fact that their observations were a combination of the objects with and without activity (that is, of a surface and then of a dust cloud), they are not comparable to the contrast displayed between the pre-impact (point source) and post-impact (extended source) observations of the Didymos-Dimorphos system. The activity of the main-belt comets is caused by the fast rotation and by the shedding of the material from the surface due to centrifugal forces, and in some of them the activity is caused by the under-surface ice \citep{MBC-review}.
So far only asteroids Scheila and P/2010 A2 have been identified as active asteroids which showed activity caused by collision \citep{Jewetal10,Jewetal11,Moreno2011}, and an impact experiment was performed on the surface of asteroid (162173) Ryugu using spacecraft Hayabusa2 \citep{Kadetal20}, but  none of these events were monitored in polarimetric mode.

The closest event that can be compared with the DART one is the outcome of the Deep Impact (DI) mission to Comet 9P/Tempel 1 (9P). DI occurred in July 2005 when a 364 kg impactor struck the surface of 9P, excavating a dust cloud of cometary material from the nucleus \citep[e.g.][]{Meechetal05}. Rather than attempting to alter the orbital path of the comet, the goal of the mission was to study the material of the ejecta. Opposite to what we observed after the DART impact on Dimorphos, the observed polarization {\it increased} by $\sim 1$\,pp as a result of the Deep Impact event, and quickly returned to the previous ‘regular’ behaviour a few days after the impact  \citep{Furetal07,Hadetal07}. The observed increase of polarization may be explained in the following terms.
Cometary particles are mainly formed by fragile aggregates of submicron sized grains \citep[e.g.][]{Horetal16} that were shattered by the impact, producing a cloud of small particles dominated by those in micron-size range \citep{Jordaetal07}. Light scattered by this type of smaller particles is expected to be more polarized than that scattered by larger particles. Furthermore, an increased amount of ice in the particles ejected from the subsurface layers may increase polarization, as shown by the comparison between the polarization from icy particles and that from silicates or organics made by \citet{MacKol22}. 
\begin{figure}
\begin{center}
\includegraphics[angle=0,width=7.0cm,trim={0.5cm 6.0cm 1.0cm 0.5cm},clip]{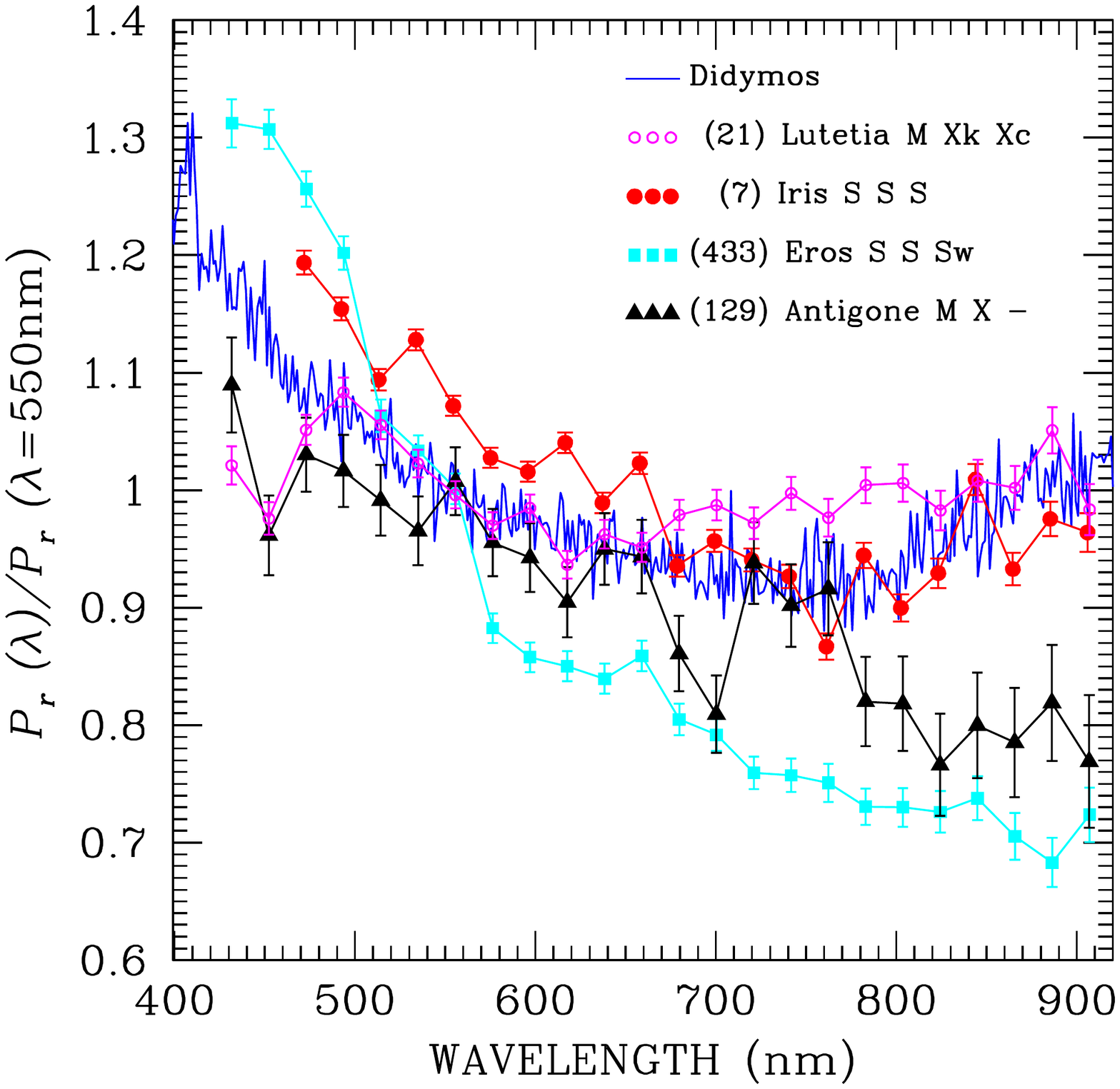}
\includegraphics[angle=0,width=7.0cm,trim={0.5cm 6.0cm 1.0cm 0.5cm},clip]{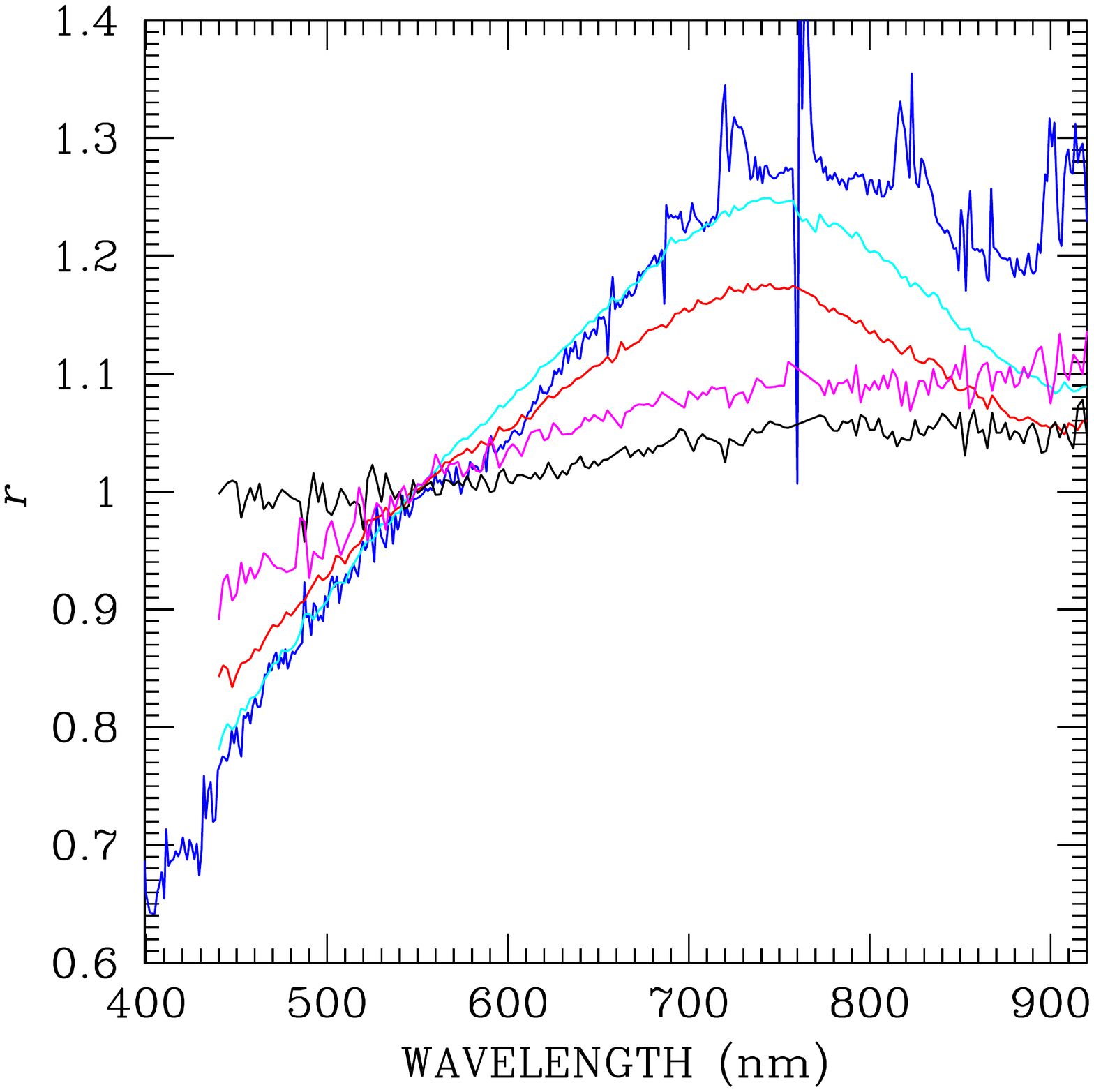}
\end{center}
\caption{\label{Fig_Comparison} Left panel: Normalised polarization spectra of Didymos obtained just before impact on 2022-09-26 compared to that of other asteroids. The right panel shows the corresponding reflectance spectra. The three symbols after asteroid name refers to the classification according to \citet{Tholen84}, \citet{BusBin02} and \citet{Demetal09}, respectively.}
\end{figure}

Our polarimetric measurements of Didymos-Dimorphos cover the three fundamental phases of the DART experiment, namely (1) the pre-impact situation; (2) a short interval of time (between 5.5\,h to $\sim 34$\,h) just following the impact; and (3) the post-impact phase over a longer timescale of $\sim 4$ weeks. Observations in imaging polarimetric mode of the latter are still ongoing and will be reported in a forthcoming paper.

The pre-impact polarimetric behaviour is typical of moderate-albedo asteroids (mostly belonging to the S taxonomic complex) observed in the positive branch of polarization. For instance, around $\alpha = 50^\circ$, S-class asteroid (433) Eros exhibits in $V$-band light a fraction of linear polarization $\sim 3$\%, that is, comparable to (but slightly less than) the values shown by the Didymos system. By contrast, low-albedo NEOs (near-Earth objects) belonging to the B and F taxonomic classes, like (101955) Bennu and (3200) Phaethon \citep{CelBennu}, exhibit a much steeper polarimetric curve. Figure~\ref{Fig_Comparison} shows that also the reflectance spectra of Didymos-Dimorphos resemble those of an S-class asteroid like (433) Eros (NEO) or the main-belt asteroid (7) Iris.  
The polarization spectrum shows an overall decrease of polarization up to wavelengths around 0.7 or 0.8\,$\mu$m,  followed by an increase that, among the objects shown in this figure, is shared only by (7) Iris. A general decrease of polarization for increasing wavelength is typical of moderate-albedo asteroids observed in the positive polarization branch \citep{Belskayaetal09,Bagetal15}.

The most obvious consequence of the impact on the polarimetric behaviour of the Didymos-Dimorphos system is a drastic decrease of the observed linear polarization. In the $B$ filter, we measure a drop by 0.9--1.0\,pp (from an expected value of $\sim 5.5$\,\% down to 4.6\,\%). The observed drop is by $\sim$ 0.5--0.6\,pp (from an expected $\sim 4.5$\,\% down to 4.0\,\%) in the $R$ filter. The relative change in polarization in the two filters was $\sim 17$ and 13\,\%, respectively. 

Our current knowledge of the phenomena of light scattering would suggest as a possible explanation the occurrence of some mechanism injecting in the system some amount of dust, provoking neutralization of polarization by multiple scattering of the light coming from the object. Another phenomenon to take into account is that, as a consequence of the space-weathering, the outer layers of asteroid surfaces become darker, i.e., more polarizing. The ejecta material, being excavated from the asteroid subsurface and not affected by the space weathering, could be lighter and produce less polarization.

The second strong evidence from our observations is that, following the impact, the shape of the polarization spectrum did not change, or changed very marginally, as shown in Figs.~\ref{Fig_Two_Spectra} and \ref{Fig_All_Spectra}. In fact,
there are much more remarkable differences between the polarization spectra of different asteroids (even of similar spectral class) shown in the left panel of Fig.~\ref{Fig_Comparison}, than between those of the system Didymos-Dimorphos observed before and after the impact.

The situation following the impact was characterized by the presence of a cloud of ejecta expanding from the Didymos satellite \citep{Li-HST-paper}, which quickly surrounded the whole binary system, before starting to slowly dissipate over longer timescales, while some fraction of this dust cloud was probably settling on the surfaces of the two companion asteroids. The dissipation of the dust cloud and the time-scale needed by the system to reach a new stable configuration will be determined by the observations that will be collected in the months to come. From the point of view of the polarimetric behaviour we have measured so far, we have to deal with the problem of a transient cloud of dust near Didymos and Dimorphos and the fact that the observed degree of linear polarization corresponds to that of the entire system of ejecta and small bodies. 

Assuming, first, that the dust cloud and the surfaces of Didymos and Dimorphos are composed of the same material, a decrease in the degree of linear polarization would follow, for example, from the realistic hypothesis that the dust cloud particles are smaller than the particles on the surfaces, thereby having higher single-scattering albedos and less pronounced positive polarization in the current phase angle range \citep{Muinonenetal96,Liu2015,Frattin2022}, while still large enough to scatter light in the geometric-optics regime \citep[$\la 50$\,$\mu$m,][and references therein]{Munetal21}.

Second, another realistic possibility would be that the ejecta particles would largely represent excavated, less space-weathered subsurface particles that would have higher single-scattering albedos than the topmost particles on the surfaces. In this second case, there would be a slight compositional gradient in the regolith particles. The polarization spectra would depend, foremost, on the single-scattering albedos and the similarity of the spectra would be explained in both cases. But we have still to infer more quantitatively the overall effect of the dust cloud on the degree of polarization, addressing carefully the contributions from Didymos, Dimorphos, and the dust cloud, and even multiple scattering among these three components \citep[see, e.g.,][]{Moreno2002}.   
The fact that the polarization spectrum did not change appreciably may be a consequence of the fact that the transient cloud of ejecta consisted of dusty material having the same optical properties of the regolith layer covering the surfaces of the two system components, assumed to be identical for the two asteroids.
This allows us to conclude that polarization behaviour with wavelength is controlled by material composition rather than bulk structure (that is, solid surface versus diffuse cloud), while the total level of polarization was strongly affected by the dispersal of material into an ejecta cloud. More detailed characterization and modelling of these effects will be the subject of a future study, but the observations already point to the DART mission as an experiment providing important clue in the understanding of how scattering material may be characterised by polarimetry.

\vspace{5mm}

\begin{small}
\noindent
Based on observations obtained at with the FORS2 instrument at the ESO Telescopes at the La Silla Paranal Observatory under program ID 109.23GL.001 and 109.23GL.002 for Didymos-Dimorphos, and 092.C-0639, 095.C-0925(A), 097.C-0853(A) for the other asteroids shown in Fig.~\ref{Fig_Comparison}. All raw data and calibrations of FORS2 are available at {\tt archive.eso.org}. 
This work was supported by the DART mission, NASA Contract No. 80MSFC20D0004 and NASA PSP grant 80NSSC21K1131. We would like to thank the great support received during observations by the VLT staff. Research by KM, AP, and MG is supported, in part, by the Academy of Finland grants No. 336546 and 345115.
\end{small}
\newpage

\appendix

\section{Observations and data reduction}\label{Appendix}
Spectropolarimetric observations were obtained with the FORS2 instrument \citep{Appetal98} attached to the Cassegrain focus of one of the ESO VLT units. FORS2 is an imaging and low-resolution spectrograph equipped with polarimetric optics, consisting of a retarder waveplate that may be rotated at fixed positions, and a Wollaston prism, following a scheme described by \citet{Appenzeller67}. Exposure times and other details of the observations are given in Table~\ref{Tab_Log}.

\subsection{Instrument setting and observing strategy}\label{Sect_Strategy}
We used grism 300V which with the order-separating filter covers a useful spectral range between approximately 430 and 920\,nm. Following the considerations by \citet{Patetal10} and the comparisons between observations obtained both with and without the order-separating filter presented by \citet{Bagetal17b}, we concluded that it was safe to use grism 300V without the order-separating filter, and extend the observed interval range into the blue, down to $\sim 400$\,nm. For all observations we set a 2" slit width, which for an extended object would provide a spectral resolving power of $\sim 220$. During the observations pre-impact, the actual spectral resolution was dictated by the seeing. In general, we were interested in the overall shape of the continuum polarization, which did not require any high spectral resolution.

CCD readout mode was set to 200\,kHz, and $2\!\times\!2$ binning, except for one case in which a slower readout (100\,kHz) mode was adopted to confirm that the contribution of readout noise---which is higher in the faster readout mode---is negligible with respect to photon-noise---at least in the context of the high S/N data needed to perform polarimetric measurements with the level of precision we were interested in. 

What we are interested in measuring is defined by Eq.~(\ref{Eq_Pr}), but the quantities $F_\perp$ and $F_\parallel$ of that equation cannot be accurately measured in such a simple way as observing through a linear polaroid without introducing large spurious effects. Various techniques and instrument designs have been proposed to maximise the accuracy of the polarimetric measurements \cite[e.g.;][]{Serkowski74,Keletal15}. The polarimetric mode of the FORS2 instrument is designed to implement the beam-swapping technique, that is, to allow the observer to swap the beams in which the polarization of the target is measured along two perpendicular directions, and recombine the signal so that the instrumental polarization is cancelled \citep[see, e.g.,][]{Bagetal09}. We obtained various observing series with a $\lambda/2$ retarder waveplate at position angles $\delta = 0^\circ$, $22.5^\circ$, $45^\circ$, \ldots, 157.5$^\circ$, or subsamples of this set, as detailed in the observing log of Table~\ref{Tab_Log}.  Reduced Stokes parameters $\pq=Q/I$\ and $\pu=U/I$\ \citep[as defined by][]{Lanetal07} were obtained using the equations of the double-difference method:
\begin{equation}
\px' = {1 \over 2 N } \sum\limits_{j=1}^N \left[ 
\left(\frac{\fo - \fe}{\fo + \fe}\right)_{\delta = \delta_j} - 
\left(\frac{\fo - \fe}{\fo + \fe}\right)_{\delta = \delta_j + 45^\circ}\right]\; ,
\label{Eq_Diff}
\end{equation}
where $\fo_{\delta}$ and $\fe_{\delta}$ are the fluxes measured in the \ord\ and \xord\ beams, respectively, with the retarder waveplate at the position angle $\delta$; for $X=Q$, $\delta_j \in\ \{ 0^\circ, 90^\circ \}$;  for $X=U$, $\delta_j \in\ \{22.5^\circ, 112.5^\circ \}$, and $N$ is the number of pairs of exposures per Stokes parameter.\footnote{We note that, in general, the symbols $f^\parallel$ and $f^\perp$ of Eq.~(\ref{Eq_Diff}) refer to fluxes measured along directions that are differently defined than those to which the fluxes $F_\parallel$ and $F_\perp$ of Eq.~(\ref{Eq_Pr}) refer.}
\begin{table}
\caption{\label{Tab_Log} Observing log. Column~4 gives the phase angle $\alpha$, col.~5 the position angle $\Phi$ of the plane containing the Sun, the Earth and the asteroid, and col.~6 the instrument position angle $\chi$ on the plane of the sky. We always tried to set $\Phi + 90^\circ \simeq \chi$, for the reasons explained in the text. Col.~7 gives the position angles $\delta$ of the $\lambda/2$ retarder waveplate used in the observing series, and the last column gives the Stokes and null parameters that have been measured in the observing series (see text).} 
\begin{center}
\begin{tabular}{ccccccll}
\hline\hline
Date       & UT    & Exp. & $\alpha$ & $\Phi$ &$\chi$&$\delta$& Observed   \\
YYYY-MM-DD & hh:mm:ss & (s)  & (deg)    & (deg)  & (deg) & (deg)         & parameters  \\
\hline
2022-08-28 & 02:49:36 &  8x300s & 23.29 &309.72&43.5&0,22.5,\ldots,157.5&\pq\&\pu\ (\nnq\&\nnu)\\  
2022-09-07 & 07:30:42 &  4x300s & 29.86 &317.31&46.0&0,45,90,135        &\pq\  (\nnq)\\    
2022-09-17 & 06:20:52 &  4x300s & 39.99 &312.54&40.0&0,45,90,135        &\pq\  (\nnq)\\   
2022-09-23 & 03:25:36 &  4x300s & 47.68 &304.38&33.0&0,45,90,135        &\pq\  (\nnq)\\   
2022-09-26 & 07:26:12 &  4x300s & 52.25 &299.10&29.0&0,45,90,135        &\pq\  (\nnq)\\ [2mm] 
2022-09-27 & 04:51:57 &  2x300s & 53.58 &297.58&27.5&0,45               &\pq\   \\ 
2022-09-28 & 09:13:50 &  8x300s & 55.34 &295.59&26.0&0,22.5,\ldots,157.5&\pq\&\pu\ (\nnq\&\nnu)\\ 
2022-09-30 & 07:53:23 &  4x200s & 58.26 &292.44&22.5&0,45,90,135        &\pq\   (\nnq)\\  
2022-10-01 & 08:14:35 &  2x250s & 59.76 &290.91&21.0&0,45               &\pq\   \\  
2022-10-24 & 06:37:24 &  4x360s & 76.16 &281.90& 12.0&0,22.5,45,67.5    &\pq\&\pu\ \\
\hline
\end{tabular}
\end{center}
\end{table}

Linear polarization was measured in the instrument reference system, that is, with respect to the direction of the principal plane of the Wollaston prism. Measurements in the instrument reference system $\px'$ were transformed into the reduced Stokes parameters $\px$ expressed with respect to a reference direction perpendicular to the scattering plane using \citep[e.g.;][]{Lanetal07}
\begin{equation}
\begin{array}{rcl}
\pq &=& \phantom{-}\pq'\cos(2\Theta) + \pu' \sin (2 \Theta) \\
\pu &=&           -\pq'\sin(2\Theta) + \pu' \cos (2 \Theta) \\
\end{array}
\label{Eq_Rota}
\end{equation}
where $\Theta = \chi + \Phi + \pi/2 + \epsilon(\lambda)$; $\chi$ is the angle of the direction of the principal plane of the Wollaston prism counted counterclockwise from the great circle passing for the target and the north celestial pole; $\Phi$ is the position angle of the scattering plane, again counted counterclockwise from the great circle passing from the object and the north celestial pole; $\epsilon(\lambda)$ is the correction to add to $\delta$ to obtain the position angle of the fast axis of the retarder waveplate measured with respect to the direction of the principal beam of the Wollaston prism; its values, which oscillates by a few degrees around zero, are tabulated in the FORS2 user manual. After this rotation, the quantity \pr\ defined in Eq.~(\ref{Eq_Pr}) is exactly given by \pq\ of Eq.~(\ref{Eq_Rota}).
To save telescope time, in most of our observations we rotated the instrument so that the position angle of the principal beam of the Wollaston prism was aligned to the direction perpendicular to the scattering plane; in other words, we set $\chi$ as to have $\Theta \simeq 0$. In that configuration, we expect $\pq' = \pq$ and (for symmetry reasons) $\pu' = \pu = 0$. The alignment was not always perfect because the exact epoch of the execution of the observations in service mode could not be predicted in advance. The observations were thus corrected assuming that $\pu = 0$ in Eq.~(\ref{Eq_Rota}) and deriving $\pq = \pq' \sec{2 \Theta}$. From the values of $\Phi$ and $\chi$ given in Table~\ref{Tab_Log} one can see that this correction was practically negligible in all cases. 

\subsection{Correction for instrumental polarization}\label{Sect_Inpol}
Previous literature \citep[e.g.;][]{PatRom06,Fosetal07,Bagetal09,Bagetal17b,Ciketal17} has highlighted the presence of a small offset value of instrumental polarization mainly due to cross-talk from $I$ to $Q$ and from $I$ to $U$, which in the centre of the field of view, and in spectropolarimetric mode, is of the order of 0.1\,\%. This was better quantified by \cite{Ciketal17} as
\begin{equation}
  \begin{array}{rcl}
    \pq^{\rm \ instr.} &=& 9.66\,\times\,10^{-8}\,\lambda + 3.29\,\times\,10^{-5}\\
    \pu^{\rm \ instr.} &=& 7.28\,\times\,10^{-8}\,\lambda - 4.54\,\times\,10^{-4}\ \; ,\\
  \end{array}
\label{Eq_Ciko}
\end{equation}
where $\lambda$ in expressed in \AA. We note that these values refer to the instrument with position angle set to 0, and they need to be rotated to take into account the instrument position angle $\chi$ at the time of the observations using formulas similar to those of Eq.~(\ref{Eq_Rota}).

\subsection{Filter polarimetry from polarization spectra}\label{Sect_BBLP}
For comparison with previous and future broadband filter measurements, we have calculated the quantities
\begin{equation}
\px(F)=\frac{\int_{0}^{\infty}\mathrm{d}{\lambda}\
\px({\lambda})\, I_X(\lambda)\, T_{\rm F}({\lambda})}{\int_{0}^{\infty}\mathrm{d}{\lambda}\ I_X(\lambda)\,T_{\rm F}({\lambda})}
\end{equation}
where $T_{F}$ is the transmission function of a certain $F$ filter, and
\begin{equation}
I_X = \frac{1}{2N} \sum_{j=1}^{N} \left[ \left(\fo + \fe\right)_{\delta = \delta_j} + \left(\fo + \fe\right)_{\delta = \delta_j + 45^\circ}\right]\; .
\end{equation}
In this work, we consider the transmission curves of the FORS2 filters $b_{\rm high}$, $v_{\rm high}$, R\_SPECIAL, and I\_BESS tabulated in the instrument web pages, and hereafter we will refer to these filters with the symbols $B$, $V$, $R$ and $I$, respectively. The actual broadband linear-polarization values do indeed depend on the filter transmission curves, but only slightly. For instance, if we were considering the transmission curve of the $V$ Bessel filter used by the ACAM instrument of the William Herschel Telescope, instead of the FORS2 $V$ filter, the value measured on September 1 would result in 5.092\,\% instead of 5.096\,\%. Effectively, these differences are negligible in the context of this study.

\subsection{Rebinning}
The polarization spectra may be rebinned to increase the S/N ratio at the expenses of the spectral resolution. We note that rebinning must be carried out on the original fluxes \fo\ and \fe\ rather than on the final reduced Stokes parameters. The dispersion of the original frames was $\sim 0.3$\,nm per pixel; all spectra shown in the Figures of this paper are shown with a $\sim 1.3$-nm wavelength bin. 

\subsection{Uncertainties}
The analytical expressions for the uncertainties of the reduced Stokes parameters may be found in App.~A of \citet{Bagetal09}. Practically, they are well approximated by the inverse of the S/N ratio accumulated in all beams used to measure the polarization in a certain wavelength bin. When fluxes are integrated to calculate the polarization in broadband filters, formally, the S/N becomes extremely high, and the real uncertainties becomes dominated by instrumental polarization. A reasonable estimate is that after the correction given by Eq.~(\ref{Eq_Ciko}), the uncertainty of the polarization measured in broadband filters is of the order of 0.05\,\% or less (see also Sect.~\ref{Sect_QC}). 

\subsection{Reflectance spectra}\label{Sect_Reflectance}
In principle, our observations could be used also to obtain reflectance spectra, by observing solar analogues at the same airmass as the target and with the same instrument setting. However, our target was observed by other instruments and telescopes to obtain exactly this information. Therefore it was decided that concentrating FORS2 on the acquisition of polarimetric data would represent a better use of telescope time. Nevertheless we used archive observations of a solar analogue, star HD\,144585, obtained with FORS2 in spectro-polarimetric mode in April 17, 2016, to calculate the approximated reflectance spectra of Didymos shown in Figs.~\ref{Fig_Two_Spectra} and \ref{Fig_Comparison}. The wavelength dependent extinction coefficients for Paranal tabulated by \citet{Patetal11} were used to correct the fluxes obtained at different airmasses. The remaining asteroid reflectance spectra of Fig.~\ref{Fig_Comparison} are from \citet{BusBin02a} and \citet{Binetal04}.

\subsection{Quality checks}\label{Sect_QC}
The quality of our observations was checked with various tests. First of all, observations obtained on August 23 and September 28 were examined to check that the polarization was all in \pq, and that \pu\ was consistent with zero. These observations were taken with $\Theta \simeq 0$, and they were also used to check that $\pq \simeq \pq'$\ as explained in Sect.~\ref{Sect_Strategy}. A small inconsistency was found in the bluest regions of the spectrum, in that the \pu\ values at shorter wavelength were found systematically slightly different from zero. We ascribed this discrepancy to a non-correct characterisation of the chromathism of the retarder waveplate $\epsilon(\lambda)$. The position angle of the polarization measured on September 28 was used to apply a small correction to the $\epsilon(\lambda)$ function tabulated in the instrument user manual---in practice decreasing by about $1^\circ$ its value around the $B$ filter. With this correction, the \pu\ values became perfectly consistent with zero also in the bluest spectral regions. It was also verified that applying the small correction of instrumental polarization given by Eq.~(\ref{Eq_Ciko}), rotated to take into account the actual instrument position angle, led to a better consistency with zero of the \pu\ profile observed on August 23. 

\begin{figure}
\begin{center}
\includegraphics[angle=0,width=7.0cm,trim={0.5cm 5.9cm 1.0cm 3.5cm},clip]{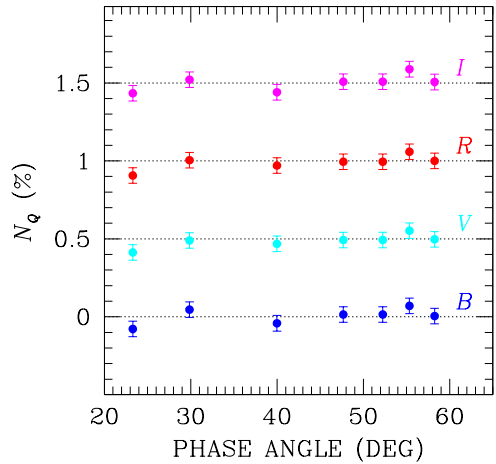}
\end{center}
\caption{\label{Fig_Null} The \nnq\ values calculated in the \emph{BVRI} filters, offset by 0, +0.5, +1.0, +1.5\,\%, respectively. Errorbars are set to 0.05\,\%. }
\end{figure}

Other important checks were systematically performed by determining the null profiles \nnq\ and \nnu\, and making sure that they were always consistent with zero within error bars. The null profiles are defined as
\begin{equation}
\nnx = {1 \over 2N} \sum\limits_{j=1}^N (-1)^{(j-1)}\,\left[ 
\left(\frac{\fo - \fe}{\fo + \fe}\right)_{\delta_j} - 
\left(\frac{\fo - \fe}{\fo + \fe}\right)_{\delta_j + 45^\circ}\right]\; .
\label{Eq_Null}
\end{equation}
and practically represent the difference between two consecutive measurements of a Stokes parameter. Instrument flexures and/or inaccurate guiding could cause the null profile to exhibit significant departures from zero, which in fact were never observed in any of our datasets. Figure~\ref{Fig_Null} shows the \nnq\ values for the phase angle at which they could be measured. All points are consistent with zero within the estimated uncertainties of 0.05\,\%.

\subsection{Spectropolarimetry of asteroids shown in Fig.~\ref{Fig_Comparison}}
Observations of asteroids (7) Iris, (21) Lutetia, (129) Antigone, and (433) Eros were obtained with the same technique, instrument, and instrument settings as for the binary system Didymos-Dimorphos. Observing dates were 
October 25, 2013 at phase angle $\alpha  =28.2^\circ$ for Iris \citep[already published by][]{Bagetal15}; 
June 2, 2015 ($\alpha=27.5^\circ$) for Lutetia;
April 17, 2015 ($\alpha=24.5^\circ$) for Antigone;
July 6, 2016 ($\alpha=28.2^\circ$) for Eros. 

\bibliography{sbabib}{}
\bibliographystyle{aasjournal}

\end{document}